\documentclass[12pt]{article}  
\usepackage{cite}
\usepackage{epsfig}
\usepackage{graphicx}
\usepackage{amsmath}
\usepackage{amssymb}
\usepackage{ulem}
\usepackage{mdwlist}          
\usepackage{color}              
\usepackage{slashed}              

\usepackage{a41}
\usepackage{color}
\usepackage[rflt]{floatflt}
\usepackage{float}                  
\usepackage{slashed}

\usepackage{array}

\usepackage[english]{babel}
\usepackage[latin1]{inputenc}
\usepackage[T1]{fontenc}
\usepackage{ae}

\usepackage{url}

\usepackage{amsmath, amsthm, amssymb}
\newtheorem{thm}{Theorem}[section]

\newtheorem{definition}[thm]{Definition}

\newcommand{\HA}{{\rm H}}

\newcommand{\Mvec}{{\rm\bf M}}

\newcommand{\ep}{\varepsilon}

\usepackage{rotating}

\usepackage{graphicx}

\newcounter{mmacnt}
\def\restartmma{\setcounter{mmacnt}{0}}
\restartmma \catcode`|=\active
\def|#1|{\mathrm{#1}}
\catcode`|=12
\newenvironment{mma}{
 \par\smallskip
 \catcode`|=\active
 \parskip=0pt\parindent=0pt 
 \small
 \def\In##1\\{%
   \def\linebreak{\hfill\break\null\qquad}%
   \refstepcounter{mmacnt}
   \hangindent=2.5em\hangafter=0
   \leavevmode
   \llap{\tiny\sffamily In[\arabic{mmacnt}]:=\kern.5em}%
   \mathversion{bold}\footnotesize$\displaystyle##1$\normalsize
   \mathversion{normal}\par
 }%
 \def\Print##1\\{%
   \def\linebreak{		\hfill\break}%
   \hangindent=2.5em\hangafter=0
   \leavevmode ##1\par}%
 \def\Out##1\\{%
   \def\linebreak{$\hfill\break\null\hfill$}%
   \kern\abovedisplayskip\par
   \hangindent=2.5em\hangafter=0
   \leavevmode
   \llap{\tiny\sffamily Out[\arabic{mmacnt}]=\kern.5em}
   \footnotesize$\displaystyle##1$\normalsize\hfill\null\par
   \kern\belowdisplayskip
 }%
 \def\Warning##1##2\\{%
   \def\linebreak{\hfill\break}%
   \hangindent=2.5em\hangafter=0
   \leavevmode
   {\scriptsize##1 : ##2}\par}%
}{%
 \par\smallskip
}

\usepackage{color}

\newenvironment{fshaded}{%
\MakeFramed {\FrameRestore}
}%
{\endMakeFramed}

\allowdisplaybreaks[4]

\begin{document}
\setlength{\baselineskip}{0.515cm}
\sloppy
\thispagestyle{empty}
\begin{flushleft}
DESY 19--118
\\
DO--TH 19/12\\
MSUHEP--19--013 \\
August 2019\\
\end{flushleft}

\mbox{}
\vspace*{\fill}
\begin{center}

{\LARGE\bf \boldmath 
The Polarized Three-Loop Anomalous}

\vspace*{3mm}
{\LARGE\bf \boldmath Dimensions from  On-Shell Massive} 

\vspace*{3mm}
{\LARGE\bf Operator Matrix Elements}

\vspace{3cm}
\large
A.~Behring$^{a,b}$
J.~Bl\"umlein$^a$, 
A.~De Freitas$^a$, 
A.~Goedicke$^{a,b,c}$,
S.~Klein$^d$,\\
A.~von Manteuffel$^e$,
C.~Schneider$^c$, and
K.~Sch\"onwald$^a$

\vspace{1.cm}
\normalsize
{\it  $^a$ Deutsches Elektronen--Synchrotron, DESY,}\\
{\it  Platanenallee 6, D-15738 Zeuthen, Germany}
\\

\vspace*{3mm}
{\it $^b$~Institut f\"ur Theoretische Teilchenphysik Campus S\"ud},\\
{\it Karlsruher Institut f\"ur Technologie (KIT) D-76128 Karlsruhe, Germany}
\\

\vspace*{3mm}
{\it $^c$~Research Institute for Symbolic Computation (RISC),\\
                          Johannes Kepler University, Altenbergerstra\ss{}e 69,
                          A-4040 Linz, Austria}\\

\vspace*{3mm}
{\it $^d$~Institut f\"ur Theoretische Teilchenphysik und Kosmologie,\\
RWTH Aachen University, D-52056 Aachen, Germany}
\\

\vspace*{3mm}
{\it $^e$~Department of Physics and Astronomy, \\ Michigan State University,
East Lansing, MI 48824, USA}


\end{center}
\normalsize
\vspace{\fill}
\begin{abstract}
\noindent
We calculate all contributions  $\propto T_F$ to the polarized three--loop anomalous dimensions in the
M--scheme using massive operator matrix elements and compare to results in the literature. This includes 
the complete anomalous dimensions $\gamma_{qq}^{(2),\rm PS}$ and $\gamma_{qg}^{(2)}$. We also obtain the 
complete two--loop polarized anomalous dimensions in an independent calculation. While for most of the 
anomalous dimensions the usual direct computation methods in Mellin $N$--space can be applied since all 
recurrences factorize at first order, this is not the case for $\gamma_{qg}^{(2)}$. Due to the necessity 
of deeper expansions of the master integrals in the dimensional parameter $\ep = D-4$, we had to
use the method of arbitrary high moments to eliminate elliptic contributions in intermediate steps.
4000 moments were generated to determine this anomalous dimension and 2640 moments turned out to be
sufficient. As an aside, we also recalculate the contributions $\propto T_F$ to the three--loop 
QCD $\beta$--function.
\end{abstract}

\vspace*{\fill}
\noindent

\newpage

\vspace*{1mm}
\noindent
\section{Introduction}
\label{sec:1}

\vspace*{1mm}
\noindent
The polarized three--loop anomalous dimensions $\gamma_{ij}^{(2)}(N)$ and splitting functions $P_{ij}^{(2)}(z)$ 
govern 
the
scale-evolution of the polarized parton distribution functions in Quantum Chromodynamics (QCD) at next-to-next-to-leading
order (NNLO) and are of importance for precision predictions at $ep$- and hadron colliders, for the analysis of the 
different fixed target experiments, for the planned electron-ion collider EIC \cite{FUTURE} and for RHIC. They are also 
instrumental for 
the measurement of the strong coupling constant $\alpha_s(M_Z)$ \cite{alphas} at these facilities and for the precise 
prediction of key processes like the polarized Drell-Yan process, jet production cross sections, and further 
processes.
With the availability of the polarized 3--loop anomalous dimensions the present next-to-leading order data analyses
of polarized deep--inelastic scattering data \cite{PDF} can be promoted to the next-to-next-to-leading order level.
Precision analyses of this kind are also relevant for the detailed study of the spin--composition of polarized
nucleons (for reviews see \cite{Lampe:1998eu,Deur:2018roz}).

A first computation of the polarized three--loop splitting functions in the M--scheme was performed in 
Ref.~\cite{Moch:2014sna}. The two--loop splitting functions have been known since 1995 
\cite{Mertig:1995ny,SP_PS1}. In the 
flavor non--singlet case, the three--loop splitting functions $P_{qq}^{(2),\rm NS-}$ are the same as in the 
unpolarized 
case \cite{Moch:2004pa} and the contributions $\propto T_F$ have been obtained in Ref.~\cite{Ablinger:2014vwa} 
as well. This 
also applies to transversity \cite{Blumlein:2009rg,Ablinger:2014vwa}. In the unpolarized case the three--loop 
splitting functions 
were calculated in Refs.~\cite{Moch:2004pa,Vogt:2004mw} and all contributions $\propto T_F$ were confirmed in 
independent 
massive calculations in Refs.~\cite{Ablinger:2010ty,
Ablinger:2014vwa,Ablinger:2014lka,Ablinger:2014nga,Ablinger:2017tan}.
Already in 2010 we have computed the odd moments $N = 1-7$ of the polarized massive OME $A_{Qg}^{(3)}$  
and $A_{qg,Q}^{(3)}$, and more recently for $N=9$. Before 2013 the corresponding moments for the massive OME 
$A_{gg,Q}^{(3)}$ were calculated, as well as a similar number of the other OMEs at three--loop order. The whole 
set of moments remained unpublished, because an important detail in the definition of the massive OMEs with massless 
external quark lines first had to be understood.

In the present paper we compute the polarized three--loop splitting functions $P_{qq}^{(2),\rm PS}$ and 
$P_{qg}^{(2)}$ and the 
parts $\propto T_F$ of the three--loop splitting functions $P_{gq}^{(2)}$ and $P_{gg}^{(2)}$ from massive 
three--loop 
operator 
matrix elements (OMEs). They are necessary for the computation of the heavy flavor contributions to deep--inelastic 
scattering in the 
region of virtualities $Q^2$ much larger than the heavy quark mass squared $m^2$. All splitting functions but $P_{qg}^{(2)}$
are calculated by applying the techniques described in Refs.~\cite{Ablinger:2015tua,Blumlein:2018cms}. In the case of
$P_{qg}^{(2)}$ we use the method of arbitrarily high Mellin moments \cite{Blumlein:2017dxp} to generate the moments of the 
$O(1/\ep)$--pole of the corresponding OME, for which a recursion is obtained by using the guessing method \cite{GUESS}.
This recurrence is finally solved by using the package {\tt Sigma} \cite{SIG1,SIG2}. 
The calculation is performed in the Larin--scheme \cite{Larin:1993tq}. As it turns out, in a massive calculation using
on--shell massive operator matrix elements special care is necessary in treating massless external fermions, as 
we will explain later. At the end of the calculation we perform a finite renormalization to the M--scheme, 
cf.~Ref.~\cite{Matiounine:1998re}, to compare to the results in Ref.~\cite{Moch:2014sna}. We would like to mention 
that the present calculation is thoroughly performed within QCD, while in Ref.~\cite{Moch:2014sna} auxiliary graviton 
interactions had to be introduced to derive the gluonic anomalous dimensions $\gamma^{(2)}_{gq}$ and 
$\gamma^{(2)}_{gg}$.

The paper is organized as follows. In Section~\ref{sec:2} we discuss the structure of the polarized unrenormalized 
three--loop massive OMEs, by which the three--loop anomalous dimensions can be calculated. If compared to the earlier
literature, an important change has been necessary for the quarkonic projector, to obtain a consistent description of 
these quantities within the Larin scheme \cite{Larin:1993tq}. The calculation methods of the master integrals are 
summarized in Section~\ref{sec:3}. The finite renormalization from the Larin to the M--scheme \cite{Matiounine:1998re}
is described in Section~\ref{sec:4}. In Section~\ref{sec:5} we present the polarized anomalous dimensions up to two 
loops,
which can be obtained from the pole contributions of $O(1/\ep^3)$ and $O(1/\ep^2)$ of the massive OMEs. In 
Section~\ref{sec:6} we present the contributions $\propto T_F$ to the three--loop anomalous dimensions, which 
are the complete anomalous dimensions for $\gamma_{qq}^{(2),\rm PS}$ and $\gamma_{qg}^{(2)}$. 
In Section~\ref{sec:7b}
we discuss the small $z$ and large $N_F$ behaviour of the splitting functions and anomalous dimensions and
Section~\ref{sec:77} contains the 
conclusions. In Appendix~\ref{sec:A} we correct two of the operator Feynman rules given in 
Ref.~\cite{Mertig:1995ny}. Appendix~\ref{sec:B} contains the splitting functions in $z$-space calculated in the 
present paper.
\section{The polarized massive Operator Matrix Elements}
\label{sec:2}

\vspace*{1mm}
\noindent
We will calculate the contributions $\propto T_F$ to the three--loop anomalous dimensions using 
unrenormalized massive operator matrix elements. Their principal structure has been given in Mellin--$N$ space in 
Ref.~\cite{Bierenbaum:2009mv}. As 
an example we consider $\hat{\hat{A}}_{Qg}^{(3)}$,
\begin{eqnarray}
   \hat{\hat{A}}_{Qg}^{(3)}&=&
                  \Bigl(\frac{\hat{m}^2}{\mu^2}\Bigr)^{3\ep/2}
                     \Biggl[
           \frac{\hat{\gamma}_{qg}^{(0)}}{6\ep^3}
             \Biggl(
                   (N_F+1)\gamma_{gq}^{(0)}\hat{\gamma}_{qg}^{(0)}
                 +\gamma_{qq}^{(0)}
                                \Bigl[
                                        \gamma_{qq}^{(0)}
                                      -2\gamma_{gg}^{(0)}
                                      -6\beta_0
                                      -8\beta_{0,Q}
                                \Bigr]
                 +8\beta_0^2
\nonumber\\ &&
                 +28\beta_{0,Q}\beta_0
                 +24\beta_{0,Q}^2
                  +\gamma_{gg}^{(0)}
                                \Bigl[
                                        \gamma_{gg}^{(0)}
                                       +6\beta_0
                                       +14\beta_{0,Q}
                                \Bigr]
             \Biggr)
          +\frac{1}{6\ep^2}
             \Biggl(
                   \hat{\gamma}_{qg}^{(1)}
                      \Bigl[
                              2\gamma_{qq}^{(0)}
                             -2\gamma_{gg}^{(0)}
                             -8\beta_0
\nonumber\\ &&
                             -10\beta_{0,Q}
                      \Bigr]
                  +\hat{\gamma}_{qg}^{(0)}
                      \Bigl[
                              \hat{\gamma}_{qq}^{(1), {\sf PS}}\{1-2N_F\}
                             +\gamma_{qq}^{(1), {\sf NS}}
                             +\hat{\gamma}_{qq}^{(1), {\sf NS}}
                             +2\hat{\gamma}_{gg}^{(1)}
                             -\gamma_{gg}^{(1)}
                             -2\beta_1
                             -2\beta_{1,Q}
                      \Bigr]
\nonumber\\
&&
                  + 6 \delta m_1^{(-1)} \hat{\gamma}_{qg}^{(0)}
                      \Bigl[
                              \gamma_{gg}^{(0)}
                             -\gamma_{qq}^{(0)}
                             +3\beta_0
                             +5\beta_{0,Q}
                      \Bigr]
             \Biggr)
          +\frac{1}{\ep}
             \Biggl(
                   \frac{\hat{\gamma}_{qg}^{(2)}}{3}
                  -N_F \frac{\hat{\tilde{\gamma}}_{qg}^{(2)}}{3}
                  +\hat{\gamma}_{qg}^{(0)}\Bigl[
                                    a_{gg,Q}^{(2)}
\nonumber\\
&&                                    -N_Fa_{Qq}^{(2),{\sf PS}}
                                          \Bigr]
                  +a_{Qg}^{(2)}
                      \Bigl[
                              \gamma_{qq}^{(0)}
                             -\gamma_{gg}^{(0)}
                             -4\beta_0
                             -4\beta_{0,Q}
                      \Bigr]
                  +\frac{\hat{\gamma}_{qg}^{(0)}\zeta_2}{16}
                      \Bigl[
                              \gamma_{gg}^{(0)} \Bigl\{
                                                        2\gamma_{qq}^{(0)}
                                                       -\gamma_{gg}^{(0)}
                                                       -6\beta_0
\nonumber\\ &&
                                                       +2\beta_{0,Q}
                                                \Bigr\}  
                             -(N_F+1)\gamma_{gq}^{(0)}\hat{\gamma}_{qg}^{(0)}
                             +\gamma_{qq}^{(0)} \Bigl\{
                                                       -\gamma_{qq}^{(0)}
                                                       +6\beta_0
                                                \Bigr\}
                             -8\beta_0^2
                             +4\beta_{0,Q}\beta_0
                             +24\beta_{0,Q}^2
                      \Bigr]
\nonumber\\ &&
                  + \frac{\delta m_1^{(-1)}}{2}
                      \Bigl[
                              -2\hat{\gamma}_{qg}^{(1)}
                              +3\delta m_1^{(-1)}\hat{\gamma}_{qg}^{(0)}
                              +2\delta m_1^{(0)}\hat{\gamma}_{qg}^{(0)}
                      \Bigr]
                  + \delta m_1^{(0)}\hat{\gamma}_{qg}^{(0)}
                       \Bigl[
                               \gamma_{gg}^{(0)}
                              -\gamma_{qq}^{(0)}
                              +2\beta_0
                              +4\beta_{0,Q}
                      \Bigr]
\nonumber\\ &&
                  -\delta m_2^{(-1)}\hat{\gamma}_{qg}^{(0)}
             \Biggr)
                 +a_{Qg}^{(3)}
                  \Biggr]. 
\label{AhhhQg3} 
\end{eqnarray}
Here we dropped the dependence on the Mellin variable $N$ from all expressions for brevity. $\hat{m}$ denotes the 
bare heavy quark mass, 
$\varepsilon = D-4$ the dimensional parameter, $\mu$ the factorization and renormalization scale, $\zeta_l, 
l \in \mathbb{N}, l \geq 2$ denote the values of the Riemann $\zeta$ function at integer argument, $N_F$ is the number of 
massless quark flavors, $\beta_i$ the expansion coefficients of the QCD $\beta$-function, $\beta_{i,Q}$ are related 
expansion 
coefficients associated to heavy quark effects, $\gamma_{ij}^{(k)}$ the expansion coefficients of the anomalous 
dimensions, and $\delta m_k^{(l)}$ the expansion coefficients of the unrenormalized quark mass. 
The above quantities depend on the color factors $C_A = N_C, C_F = (N_C^2-1)/(2 N_C), T_F = 1/2$ for $SU(N_C)$ and 
$N_C= 3$ for QCD, cf. e.g. Ref.~\cite{Bierenbaum:2009mv}. The coefficients $a_{ij}^{(k)}$ 
denote the constant terms of the OMEs at $k$--loop order and $\bar{a}_{ij}^{(k)}$ the corresponding terms 
at $O(\varepsilon)$, 
cf.~\cite{Buza:1996xr,Klein:2009ig,Blumlein:2019zux,Hasselhuhn:2013swa,POL19,PVFNS}.\footnote{The 
structure of the unrenormalized OMEs up to 3--loop orders given in Ref.~\cite{Bierenbaum:2009mv}
partly refers to earlier work by van Neerven et al. in the unpolarized 
case~\cite{Buza:1995ie,Buza:1996wv} to allow for a more direct comparison. This particularly 
concerns the definition of $a_{Qg}^{(2)}$ and $a_{gg,Q}^{(2)}$.}
In the constant terms $O(\ep^0)$ also multiple zeta values \cite{Blumlein:2009cf}
contribute at fixed values of $N$.
Furthermore, we use the convention
\begin{eqnarray}\label{eq:r1}
\hat{f}(N_F)   &=&  f(N_F+1) - f(N_F)   \\
\label{eq:r2}
\tilde{f}(N_F) &=&  \frac{\displaystyle f(N_F)}{\displaystyle N_F}~.
\end{eqnarray}
The OME $\hat{\hat{A}}_{Qg}^{(3)}$ consists out of the one--particle irreducible part and a  reducible part.
These are related by
\begin{eqnarray}
\hat{\hat{A}}_{Qg,\rm irr}^{(2)} &=& \hat{\hat{A}}_{Qg}^{(2)} +\hat{\hat{A}}_{Qg}^{(1)} 
\hat{\Pi}_{(1)}\left(\frac{m^2}{\mu^2}\right),\\ 
\hat{\hat{A}}_{Qg}^{(3)} &=& \hat{\hat{A}}_{Qg}^{(3),\rm irr}
                                    +\hat{\hat{A}}_{Qg,\rm irr}^{(2)} \hat{\Pi}^{(1)}\left(\frac{m^2}{\mu^2}\right)
                                    -\hat{\hat{A}}_{Qg}^{(1)} \hat{\Pi}^{(2)}\left(\frac{m^2}{\mu^2}\right) 
                                    - \hat{\hat{A}}_{Qg}^{(1)} 
\left[\hat{\Pi}^{(1)}\left(\frac{m^2}{\mu^2}\right)\right]^2,
\end{eqnarray}
where $\hat{\Pi}^{(k)}\left(\frac{m^2}{\mu^2}\right)$ denote the expansion coefficients of the massive 
contributions to the gluon vacuum polarization, see.~\cite{Bierenbaum:2009mv}. The local operator insertions
can be resummed into propagator--like structures, by virtue of an auxiliary parameter $x$, which form  
corresponding generating functions, cf.~\cite{Ablinger:2014vwa}.

In total there are six massive OMEs, $A_{qq,Q}^{(3),\rm PS}, A_{Qq}^{(3),\rm PS}, A_{qg,Q}^{(3)}, A_{Qg}^{(3)}, 
A_{gq,Q}^{(3)}$ and $A_{gg,Q}^{(3)}$, which we calculate in the polarized case. Furthermore, there is the 
non--singlet OME 
$A_{qq,Q}^{(3),\rm NS}$ which has already been calculated in Ref.~\cite{Ablinger:2014vwa}. The latter quantity, due 
to a known 
Ward identity, can be given in the $\overline{\sf MS}$ scheme. From the poles $O(1/\ep^3)$ one 
obtains the 
one--loop anomalous dimensions and from the poles $O(1/\ep^2)$ the complete two--loop anomalous dimensions, 
while the 
contributions $\propto T_F$ to the three--loop anomalous dimensions are extracted from the pole terms of $O(1/\ep)$.

To describe $\gamma^5$ in $D = 4 + \ep$-dimensions, we work in the Larin scheme \cite{Larin:1993tq}\footnote{
For other schemes see Refs.~\cite{HVBM}. For a discussion of the necessary finite renormalizations see \cite{FINREN}.} 
and express $\gamma^5$ by
\begin{eqnarray}
\gamma^5 &=& \frac{i}{24} \ep_{\mu \nu \rho \sigma} \gamma^\mu \gamma^\nu \gamma^\rho 
\gamma^\sigma,
\\
\Delta \hspace*{-2.5mm} \slash ~\gamma^5 &=& \frac{i}{6} \ep_{\mu \nu \rho \sigma} \Delta^\mu \gamma^\nu \gamma^\rho 
\gamma^\sigma.
\end{eqnarray}
The Levi-Civita symbols are now contracted in $D$ dimensions,
\begin{eqnarray}
\label{eq:epcontr}
\ep_{\mu \nu \rho \sigma}  \ep^{\alpha \lambda  \tau  \gamma} = - {\rm Det}[g_\omega^\beta], 
~~~~\beta = \alpha, \lambda,  \tau,  \gamma;~~\omega = \mu, \nu, \rho, \sigma.
\end{eqnarray}
In the calculation of the OMEs with external on--shell gluonic and quarkonic states 
we use the projectors  $P_g$ and $P_q$ for the amplitudes $\hat{G}^{ab}_{\mu\nu}$
and $\hat{G}_l^{ij}$. External 
ghost states do not contribute in the polarized states to three--loop order since the 
corresponding traces turn out to vanish.
The gluonic projector is given by
\begin{eqnarray}
\label{eq:Pg}
P_g \hat{G}^{ab}_{\mu\nu} = \frac{\delta^{ab}}{N_C^2-1} \frac{1}{(D-2)(D-3)} (\Delta p)^{-N-1} 
\ep^{\mu\nu\rho\sigma} 
\Delta_\rho p_\sigma \hat{G}^{ab}_{\mu\nu}.
\end{eqnarray}
The following quarkonic projector has been proposed in Ref.~\cite{Buza:1996xr}
\begin{eqnarray}
\label{eq:Pq}
\check{P}_q \hat{G}_l^{ij} = -\delta_{ij} \frac{i  (\Delta.p)^{-N}}{16 N_C (D-1)(D-2)(D-3)} \ep_{\mu \nu \rho \sigma} 
{\rm tr} \left[p 
\hspace*{-2mm} 
\slash 
\gamma^\mu \gamma^\nu \gamma^\rho \gamma^\sigma \hat{G}_l^{ij}\right].
\end{eqnarray}
Using it and (\ref{eq:epcontr}) will imply the necessity of a finite renormalization of
the two--loop anomalous dimensions $\gamma_{qq}^{(1), \rm PS}$ \cite{Bierenbaum:2007pn} and 
$\gamma_{gq}^{(1)}$ \cite{Hasselhuhn:2013swa}, the structure and occurrence of which we did not understand for 
a series of years. Already in \cite{Bierenbaum:2007pn,Klein:2009ig} and an early version of \cite{POL19} we 
reanalyzed the 
pure singlet case at two--loop order to get agreement with the result of 
\cite{Buza:1996xr}, which has been given there without presenting details. In the recent complete 
analytic calculation of the polarized massive two--loop Wilson coefficient in the whole kinematic 
range \cite{Blumlein:2019zux} it turned out that $\gamma_{qq}^{(1), \rm PS}$ did not receive
a finite renormalization, as has been observed in the calculation of the polarized massless 
Wilson coefficient in \cite{Zijlstra:1993sh,Vogt:2008yw,CODE} before.\footnote{Note that different schemes have been
used in Ref.~\cite{Zijlstra:1993sh}.} 

For external quarkonic states a modified treatment compared to (\ref{eq:Pq}) therefore has to be applied.
In the limit of a vanishing external light quark mass two bi--spinor structures survive, see \cite{POL19}. They can 
be mapped to the following projector
\begin{eqnarray}
\label{eq:PqNEW}
P_q \hat{G}_l^{ij} = - \delta_{ij} \frac{i (\Delta.p)^{-N-1}}{4 N_C (D-2)(D-3)} \ep_{\mu \nu p \Delta} {\rm tr} \left[p 
\hspace*{-2mm} 
\slash 
\gamma^\mu \gamma^\nu \hat{G}_l^{ij}\right]
\end{eqnarray}
which yields the proper definition in the Larin scheme in the case of the massive OMEs with massless 
external quark lines, unlike 
Eq.~(\ref{eq:Pq}). The existence of a single 
projector (\ref{eq:PqNEW}) is of advantage since the calculation techniques described
below only had to be modified minimally if compared to the unpolarized case. This
also applies to the calculation of a series of fixed moments using {\tt MATAD} 
\cite{Steinhauser:2000ry}. 

The Feynman diagrams contributing to the massive OMEs were generated by the code {\tt QGRAF} 
\cite{{Nogueira:1991ex}}\footnote{See Ref.~\cite{Bierenbaum:2009mv} for the implementation of 
the local operators.}. The Dirac algebra has been performed using {\tt FORM} \cite{FORM} and
the color configurations were calculated using the package {\tt Color}  
\cite{vanRitbergen:1998pn}. The Feynman integrals were reduced to master integrals using the 
integration-by-parts (IBP) relations \cite{IBP} implemented in  
the package {\tt Reduze 2} \cite{Studerus:2009ye,vonManteuffel:2012np}.\footnote{The 
package {\tt Reduze 2} uses the packages {\tt FERMAT} \cite{FERMAT} and {\tt Ginac} 
\cite{Bauer:2000cp}.}. There are different techniques available to calculate the master 
integrals, cf.~Refs.~\cite{Blumlein:2017dxp,Ablinger:2015tua}, which we discuss in the next section.

The constant contributions to the two--loop OMEs $a_{ij}^{(2)}$ in the Larin scheme are given in 
\cite{POL19} for $a_{Qg}^{(2)}$, \cite{POL19,Blumlein:2019zux} for $a_{qq,Q}^{(2),\rm PS}$,
\cite{Hasselhuhn:2013swa} for $a_{gg,Q}^{(2)}$, \cite{PVFNS} for $a_{gq,Q}^{(2)}$.
In the non--singlet case we obtain
\begin{eqnarray}
a_{qq,Q}^{(2),\rm NS} &=& \textcolor{blue}{C_F T_F} \Biggl\{ 
\frac{R_1}{54 N^3 (N+1)^3} 
+ \Biggl(\frac{2 (2 + 3 N + 3 N^2)}{3 N ( N+1)} - \frac{8}{3} S_1 \Biggr) \zeta_2 - 
    \frac{224}{27} S_1 + \frac{40}{9} S_2 - \frac{8}{3} S_3 \Biggr\}
\nonumber\\
\\
\overline{a}_{qq,Q}^{(2),\rm NS} &=& \textcolor{blue}{C_F T_F} \Biggl\{ 
\frac{R_2}{648 N^4 (1 + N)^4} + 
      \Biggl(\frac{2 (2 + 3 N + 3 N^2)}{9 N ( N+1)} - \frac{8}{9} S_1 \Biggr) \zeta_3
    + \Biggl(\frac{R_3}{18 N^2 (N+1)^2} 
\nonumber\\ &&
- 
       \frac{20}{9} S_1 
+ \frac{4}{3} S_2\Biggr) \zeta_2 
- \frac{656}{81} S_1 + \frac{112}{27} S_2 - \frac{20}{9} S_3 + \frac{4}{3} S_4
\Biggr\},
\\
R_1 &=& 72 + 240 N + 344 N^2 + 379 N^3 + 713 N^4 + 657 N^5 + 219 N^6,
\\
R_2 &=& -432 - 1872 N - 3504 N^2 - 3280 N^3 + 1407 N^4 + 7500 N^5 + 9962 N^6 + 6204 N^7 
\nonumber\\ &&
+ 1551 N^8,
\\
R_3 &=& -12 - 28 N - N^2 + 6 N^3 + 3 N^4.
\end{eqnarray}
Here $\overline{a}_{qq,Q}^{(2),\rm NS}$ denotes the $O(\ep)$ contribution needed in the renormalization of the 
three--loop OME. We note that the use of the projectors (\ref{eq:Pg}, \ref{eq:PqNEW}) for the massive OMEs allow 
to extract the contributions to the anomalous dimensions in the  Larin--scheme\footnote{The representation of the 
two--loop massive OMEs contributing to the structure function $g_1$ in the M--scheme are presented in 
Ref.~\cite{POL19}.} from all pole terms of $O(\ep^{-k}),~~k=3,2,1$. Their finite renormalization to translate to the 
M--scheme is described in Section~\ref{sec:4}.
\section{The calculation methods}
\label{sec:3}

\vspace*{1mm}
\noindent
For the pole terms of the OMEs $A_{qq,Q}^{(3),\rm PS}, A_{Qq}^{(3),\rm PS}, A_{qg,Q}^{(2)}, A_{gq,Q}^{(3)}$ and
$A_{gg,Q}^{(3)}$ the contributing master integrals can be calculated by the standard techniques such as 
the method of hypergeometric functions \cite{HYP,SLATER}, the method of hyperlogarithms
\cite{Brown:2008um,Ablinger:2014yaa,Panzer:2014caa}, the solution of ordinary differential equation systems 
\cite{DEQ,Ablinger:2015tua,Ablinger:2018zwz} and the Almkvist--Zeilberger algorithm \cite{AZ,Ablinger:PhDThesis}, 
being used in a combination, since in higher order in the dimensional parameter no elliptic integrals 
contribute.\footnote{For a recent survey on these methods see \cite{Blumlein:2018cms}.} Some of the simpler integrals 
have been calculated using Mellin--Barnes representations and using the codes in \cite{MB}. Most of the master 
integrals were already available from the calculation 
of the unpolarized three--loop anomalous dimensions in Ref.~\cite{Ablinger:2017tan}. Only in a few cases some further 
differential equations had to be solved to obtain all master integrals. In all of the above methods corresponding sum 
representations have been derived which were solved using the difference--field techniques 
\cite{Karr:81,Schneider:01,Schneider:05a,Schneider:07d,Schneider:10b,Schneider:10c,Schneider:15a,Schneider:08c} 
of the packages {\tt Sigma} \cite{SIG1,SIG2}, {\tt EvaluateMultiSums}, {\tt SumProduction} \cite{EMSSP}, and using
{\tt HarmonicSums} \cite{Vermaseren:1998uu,Blumlein:1998if,HARMONICSUMS,Ablinger:PhDThesis,Ablinger:2011te,Ablinger:2013cf,
Ablinger:2014bra}.

This is, however, different for $A_{Qg}^{(3)}$. Due to the structure of the IBP relations some higher expansion in 
$\ep$ is 
necessary also to extract the term $\propto 1/\ep$. Here one would encounter elliptic contributions 
\cite{Ablinger:2017bjx,Blumlein:2018aeq} by using the above techniques. We therefore apply the method of arbitrarily 
large moments \cite{Blumlein:2017dxp} in this case.\footnote{This method has been successfully applied also in a series 
of other calculations, cf.~\cite{Ablinger:2017tan,GUESS1}.} Here one works in moment--space and the IBP relations
are expressed in terms of recurrences for the master integrals. Using these relations one generates 
systematically higher and higher moments both for the master integrals and the operator matrix elements.

The projection to the analytic representation of the moments of the 
master integrals allows to treat also elliptic and higher structures. Finally one obtains the moments of the OME.
They are used to derive a difference equation by the method of guessing \cite{GUESS}\footnote{For an early application 
to large systems in Quantum Field Theory see Ref.~\cite{Blumlein:2009tj}.} implemented in {\tt Sage}
\cite{SAGE,GSAGE}, based on very fast integer algorithms. 
We generated 2000 Mellin moments, which allowed to find most of the recurrences for all seventeen color--$\zeta$ 
projections. To determine the recurrences of the projections $C_F C_A T_F$ and $C_A^2 T_F$ we used 4000 moments, out 
of which 2640 turned out to be sufficient. 
Here we refer to representations in terms of even and odd moments, with 
the even moments being unphysical. The analytic continuation is finally performed from the odd moments only.
The characteristics of the recurrences for the different color--$\zeta$ factors contributing to the $1/\ep$ term of 
the unrenormalized massive OME $A_{Qg}^{(3)}$ are summarized in Table~\ref{TAB1}. For all the pole terms these 
recurrences are first--order factorizable and can be solved by applying the package {\tt Sigma}.
Here some color--$\zeta$ structures contribute for technical reasons, which cancel in the final expression.

All anomalous dimensions can be expressed by nested harmonic sums \cite{Vermaseren:1998uu,Blumlein:1998if}
\begin{eqnarray}
S_{b,\vec{a}}(N) &=& \sum_{k=1}^N \frac{({\rm sign}(b))^k}{k^{|b|}} S_{\vec{a}}(k),~~~S_\emptyset = 1~~~,b, a_i \in
\mathbb{Z} \backslash \{0\}. 
\end{eqnarray}
To provide comparisons on a diagram-by-diagram basis we have calculated the first few Mellin moments for $N = 1, 3, 
5, 7, 9$
using {\tt MATAD} \cite{Steinhauser:2000ry}.
\begin{table}[H]\centering
\def\arraystretch{1.5}%
\begin{tabular}{|l|r|r|}
\hline
\multicolumn{1}{|c}{color/$\zeta$} & \multicolumn{1}{|c}{order} & \multicolumn{1}{|c|}{degree}\\  
\hline
$C_F T_F^2$                &  7  &  68    \\
$C_F T_F^2 \zeta_2$        &  3  &  17    \\
$C_F T_F^2 N_F$            &  7  &  68    \\
$C_F T_F^2 N_F \zeta_2$    &  3  &  17    \\
$C_F^2 T_F$                & 22  & 283    \\
$C_F^2 T_F \zeta_2$        &  6  &  32    \\
$C_F^2 T_F \zeta_3$        &  2  &  10    \\
$C_A T_F^2$                & 10  &  85    \\
$C_A T_F^2 \zeta_2$        &  3  &  12    \\
$C_A T_F^2 N_F$            & 14  & 131    \\
$C_A T_F^2 N_F \zeta_2$    &  4  &  16    \\
$C_F C_A T_F$              & 30  & 484    \\
$C_F C_A T_F \zeta_2$      &  8  &  46    \\
$C_F C_A T_F \zeta_3$      &  3  &  19    \\
$C_A^2 T_F$                & 30  & 472    \\
$C_A^2 T_F \zeta_2$        & 10  &  57    \\
$C_A^2 T_F \zeta_3$        &  4  &  19    \\
\hline
\end{tabular}
\def\arraystretch{1}%
\caption[]{\label{TAB1}
\sf Characteristics of the recurrences contributing to the anomalous dimension $\gamma_{qg}^{(2)}$.}
\end{table}
\section{The finite renormalizations from the Larin to the \newline M--scheme}
\label{sec:4}

\vspace*{1mm}
\noindent
We would like to compare to the results obtained in Ref.~\cite{Moch:2014sna} which are given in the
M--scheme. This scheme was defined in implicit form in Ref.~\cite{Matiounine:1998re}. Up to two--loop order 
it is the same as the one in which the results of Refs.~\cite{Mertig:1995ny,SP_PS1} were obtained.
The anomalous dimensions have the expansions in the non--singlet and singlet case
\begin{eqnarray}
\gamma_{qq}^{\rm NS, M} &=& \sum_{k=0}^\infty a_s^{k+1} \gamma_{qq}^{(k),\rm NS, M}
\\ 
\gamma_{ij}^{\rm M} &=& \sum_{k=0}^\infty a_s^{k+1} \gamma_{ij}^{(k),\rm  M},~~i,j \in \{q,g\}.
\end{eqnarray}
At leading order, the anomalous dimensions are scheme--invariant.
The finite renormalizations between the Larin and the M--scheme to three--loop order can be obtained following 
\cite{Matiounine:1998re}, see also \cite{Moch:2014sna}, and are given by:
\begin{eqnarray}
        \gamma_{qq}^{(1),\text{NS},\rm M} &=& \gamma_{qq}^{(1),\text{NS},\rm L} + 2 \beta_0 
z_{qq}^{(1)},
\\
        \gamma_{qq}^{(1),\text{PS}, \rm M} &=& \gamma_{qq}^{(1),\text{PS}, \rm L},
\\
        \gamma_{qg}^{(1),\rm M} &=& \gamma_{qg}^{(1), \rm L} + \gamma_{qg}^{(0)} z_{qq}^{(1)},
\\
        \gamma_{gq}^{(1),\rm M} &=& \gamma_{gq}^{(1), \rm L} - \gamma_{gq}^{(0)} z_{qq}^{(1)},
\\
        \gamma_{gg}^{(1),\rm M} &=& \gamma_{gg}^{(1), \rm L}.
\\
       \gamma_{qq}^{(2),\text{NS}, \rm M} &=& \gamma_{qq}^{(2),\text{NS}, \rm L} - 2 \beta_0 \left( 
\bigl( 
z_{qq}^{(1)} \bigr)^2 - 2 z_{qq}^{(2),\text{NS}} \right) + 2 \beta_1 z_{qq}^{(1)} ,
\\
        \gamma_{qq}^{(2), \rm PS,M} &=& \gamma_{qq}^{(2), \rm PS,L} + 4 \beta_0 z_{qq}^{(2),\rm PS},
\\
        \gamma_{qg}^{(2), \rm M} &=& \gamma_{qg}^{(2), \rm L} + \gamma_{qg}^{(1), \rm M} 
z_{qq}^{(1)} + 
\gamma_{qg}^{(0)} \left( z_{qq}^{(2)} - \bigl( z_{qq}^{(1)} \bigr)^2 \right),
\\
        \gamma_{gq}^{(2), \rm M} &=& \gamma_{gq}^{(2),\rm L} - \gamma_{gq}^{(1),\rm M} z_{qq}^{(1)} 
- 
\gamma_{gq}^{(0)} z_{qq}^{(2)} ,
\\
        \gamma_{gg}^{(2),\rm M} &=& \gamma_{gg}^{(2),\rm L} ,
\end{eqnarray}
with \cite{Matiounine:1998re}
\begin{eqnarray}
        z_{qq}^{(1)} &=& 
- \frac{8 
\textcolor{blue}{C_F}}{N(N+1)},
\\
z_{qq}^{(2),\text{NS}} 
&=&
\textcolor{blue}{C_F T_F N_F} 
                \frac{16 \big(-3-N+5 N^2\big)}{9 N^2 (1+N)^2}
+ \textcolor{blue}{C_A C_F} \Biggl\{
        -\frac{4 Q_1}{9 N^3 (1+N)^3}
        -\frac{16}{N (1+N)} S_{-2}
\Biggr\}
\nonumber \\ &&
+ 
\textcolor{blue}{C_F^2} 
\Biggl\{
          \frac{8\big(2+5 N+8 N^2+N^3+2 N^4\big)}{N^3 (1+N)^3} 
        + \frac{16(1+2 N)}{N^2 (1+N)^2} S_1
\nonumber \\ &&
        + \frac{16}{N (1+N)} S_2 
        + \frac{32}{N (1+N)} S_{-2}
\Biggr\},
\\
z_{qq}^{(2),\text{PS}} 
&=& 8 
\textcolor{blue}{C_F T_F N_F} 
\frac{(N+2)(1+N-N^2)}{N^3(N+1)^3},
\\
        z_{qq}^{(2)} &=& 
z_{qq}^{(2),\text{NS}} + 
z_{qq}^{(2),\text{PS}} 
\end{eqnarray}
and
\begin{eqnarray}
Q_1 =  103 N^4 + 140 N^3 + 58 N^2 + 21 N + 36.
\end{eqnarray}
Specifically one obtains the following transformations:
\begin{eqnarray}
        \hat{\gamma}_{qq}^{(1),\text{NS}, \rm M} &=& \hat{\gamma}_{qq}^{(1),\text{NS}, \rm L} + 
\textcolor{blue}{C_F T_F N_F} \frac{64}{3N(N+1)}, 
\\
        \gamma_{qg}^{(1),\rm M} &=& \gamma_{qg}^{(1),\rm L} + \textcolor{blue}{C_F T_F N_F} 
\frac{64(N-1)}{N^2(N+1)^2},
\\
        \hat{\gamma}_{gq}^{(1), \rm M} &=& \hat{\gamma}_{gq}^{(1), \rm L}, 
\\
        \hat{\gamma}_{qq}^{(2),\text{NS},\rm M} &=& \hat{\gamma}_{qq}^{(2),\text{NS}, \rm L}
- \textcolor{blue}{C_F T_F^2(2 N_F+1)} \frac{256 \big(-3-N+5 N^2\big)}{27 N^2 (1+N)^2}
+ \textcolor{blue}{C_A C_F T_F}
\Biggl\{
                \frac{64 T_1}{27 N^3 (1+N)^3}
\nonumber \\ &&
                +\frac{256}{3 N (1+N)} S_{-2}
\Biggr\}
+ \textcolor{blue}{C_F^2 T_F}
\Biggl\{
                -\frac{64}{3 N^3 (1+N)^3} \big(4+2 N
                +5 N^2-4 N^3+N^4\big)
\nonumber \\ &&
                -\frac{256 (1+2 N)}{3 N^2 (1+N)^2} S_1
                -\frac{256}{3 N (1+N)} S_2
                -\frac{512}{3 N (1+N)} S_{-2}
\Biggr\},
\\
        \gamma_{qq}^{(2),\text{PS},\rm M} &=& \gamma_{qq}^{(2),\text{PS}, \rm L}
 -\frac{(N+2)(1+N-N^2)}{3 N^3(N+1)^3} \left[128 \textcolor{blue}{C_F T_F^2 N_F^2} - 352 \textcolor{blue}{C_A C_F T_F 
N_F}\right],  
\\
        \gamma_{qg}^{(2),\rm M} &=& \gamma_{qg}^{(2), \rm L}
- \textcolor{blue}{C_F T_F^2 N_F^2} \frac{64 (N-1)\big(18+21 N-17 N^2-N^3+10 N^4\big)}{9 N^4 
(1+N)^4}
\nonumber \\ &&
+ \textcolor{blue}{C_A C_F T_F N_F}
\Biggl\{
                \frac{32 T_2}{9 N^4 (1+N)^4}
                +\frac{512}{N^2 (1+N)^3} S_1
                -\frac{128 (N-1)}{N^2 (1+N)^2} \Bigl( S_1^2 + S_2 + S_{-2} \Bigr)
\Biggr\}
\nonumber \\ &&
+ \textcolor{blue}{C_F^2 T_F N_F}   
\Biggl\{
                \frac{64 (N-1) \big(2+9 N^2+3 N^3\big)}{N^3 (1+N)^4}
                -\frac{128 (N-1) (3+4 N)}{N^3 (1+N)^3} S_1
\nonumber \\ &&
                +\frac{128 (N-1)}{N^2 (1+N)^2} \Bigl( S_1^2 - 2 S_2 - 2 S_{-2} \Bigr)
\Biggr\},
\\
        \hat{\gamma}_{gq}^{(2),\rm M} &=& \hat{\gamma}_{gq}^{(2), \rm L}
+ \textcolor{blue}{C_F^2 T_F}
\Biggl\{
        \frac{32 (2+N) (6+5 N) \big(3+N-N^2+10 N^3\big)}{9 N^4 (1+N)^4}
        -\frac{256 (2+N)}{3 N^2 (1+N)^2} S_1 
\Biggr\},
\nonumber\\ 
\end{eqnarray}
with the polynomials
\begin{eqnarray}
        T_1 &=& 36-12 N+59 N^2+274 N^3+203 N^4 ,
\\
        T_2 &=& -108-237 N+71 N^2-226 N^3+73 N^4+139 N^5.
\end{eqnarray}
A priori, it has not been clear whether the use of the Larin scheme in
the massive case leads to results which are equivalent to the HVBM
scheme~\cite{HVBM}, which is known to occur in the massless case, cf.~\cite{Moch:2014sna}. 
For the calculation of the anomalous dimensions it turns out that this is indeed the case.
For the future it still remains to 
analyze all conditions implied by the Slavnov--Taylor identities, which are violated by dimensional 
regularization in both the Larin and HVBM schemes \cite{Larin:1993tq,HVBM}, in calculations of anomalous dimensions 
and Wilson coefficients from two--loop order onward. 
\section{The polarized anomalous dimensions up to two--loop order}
\label{sec:5}

\vspace*{1mm}
\noindent
Here and in the following we reduce the representations in Mellin--$N$ space to bases by applying their algebraic 
relations, 
cf.~\cite{Blumlein:2003gb}. We will also use the shorthand notation $S_{\vec{a}}(N) \equiv S_{\vec{a}}$.
The leading order anomalous dimensions are given by
\begin{eqnarray}
\gamma_{qq}^{(0)} &=& \textcolor{blue}{C_F} \Biggl\{-\frac{2 (2 + 3 N + 3 N^2)}{N (N+1)} + 8 S_1\Biggr\}
\\
\gamma_{qg}^{(0)} &=& -\textcolor{blue}{T_F N_F} \frac{8 (N-1)}{N (N+1)}
\\
\gamma_{gq}^{(0)} &=& -\textcolor{blue}{C_F} \frac{4 (2 + N)}{N (N+1)}
\\
\gamma_{gg}^{(0)} &=& \textcolor{blue}{T_F N_F} \frac{8}{3} +
\textcolor{blue}{C_A} \Biggl\{- \frac{2 (24 + 11 N + 11 N^2)}{3 N (1 + N)} + 8 S_1\Biggr\}
\end{eqnarray}
They are scheme--independent and agree with the results given in Refs.~\cite{Gross:1973ju,Georgi:1951sr,Sasaki:1975hk,
Ahmed:1975tj,Altarelli:1977zs}\footnote{The foregoing paper \cite{Ito:1975pf} was not fully correct, see also 
\cite{Blumlein:2012bf} for a survey on earlier work.}.

The next-to-leading order anomalous dimensions are given by \cite{Floratos:1977au,Curci:1980uw,GonzalezArroyo:1979df,
Mertig:1995ny,SP_PS1,Moch:1999eb,Vogt:2008yw,Moch:2014sna,Blumlein:2019zux,POL19,PVFNS}
\begin{eqnarray}
\gamma_{qq}^{(1),\rm NS} &=& 
\textcolor{blue}{C_F} \Biggl\{
        \textcolor{blue}{T_F N_F} \Biggl[
                \frac{4 P_1}{9 N^2 (1+N)^2}
                -\frac{160}{9} S_1
                +\frac{32}{3} S_2
        \Biggr]
        +\textcolor{blue}{C_A} \Biggl[
                \frac{P_2}{9 N^3 (1+N)^3}
                +\frac{536}{9} S_1
                -\frac{88}{3} S_2
\nonumber\\ &&
                +16 S_3
                +\Biggl(
                        -\frac{16}{N (1+N)}
                        +32 S_1
                \Biggr) S_{-2}
                +16 S_{-3}
                -32 S_{-2,1}
        \Biggr]
\Biggr\}
\nonumber\\ &&
+\textcolor{blue}{C_F^2} \Biggl\{
        \frac{P_3}{N^3 (1+N)^3}
        +\Biggl(
                \frac{16 (1+2 N)}{N^2 (1+N)^2}
                -32 S_2
        \Biggr) S_1
        +\frac{8 \big(
                2+3 N+3 N^2\big)}{N (1+N)} S_2
\nonumber\\ &&
        -32 S_3
        +\Biggl(
                \frac{32}{N (1+N)}
                -64 S_1
        \Biggr) S_{-2}
        -32 S_{-3}
        +64 S_{-2,1}
\Biggr\},
\\
P_1 &=& 3 N^4+6 N^3+47 N^2+20 N-12,
\\
P_2 &=& -51 N^6-153 N^5-757 N^4-995 N^3-496 N^2-156 N-144,
\\
P_3 &=& -3 N^6-9 N^5-9 N^4-27 N^3+24 N^2+32 N+24,
\\
\gamma_{qq}^{(1),\rm PS} &=&
\textcolor{blue}{C_F T_F N_F}
\frac{16 (2+N) \big(
        1+2 N+N^3\big)}{N^3 (1+N)^3},
\\
\gamma_{qg}^{(1)} &=&
\textcolor{blue}{C_F T_F N_F} \Biggl\{
        -\frac{8 (N-1) \big(
                2-N+10 N^3+5 N^4\big)}{N^3 (N+1)^3}
        +\frac{32 (N-1)}{N^2 (N+1)} S_1
        -\frac{16 (N-1)}{N (N+1)} S_1^2
\nonumber\\ &&
        +\frac{16 (N-1)}{N (N+1)}   S_2
\Biggr\}
+ \textcolor{blue}{C_A T_F N_F} \Biggl\{
        -\frac{16 P_4}{N^3 (N+1)^3}
        -\frac{64}{N (N+1)^2} S_1
        +\frac{16 (N-1)}{N (1+N)} S_1^2
\nonumber\\ &&
        +\frac{16 (N-1)}{N (1+N)} S_2
        +\frac{32 (N-1)}{N (1+N)} S_{-2}
\Biggr\},
\\
P_4 &=& N^5+N^4-4 N^3+3 N^2-7 N-2,
\\
\gamma_{gq}^{(1)} &=&
\textcolor{blue}{C_F} \Biggl\{
        \textcolor{blue}{T_F N_F}  \Biggl[
                \frac{32 (2+N) (2+5 N)}{9 N (N+1)^2}
                -\frac{32 (2+N)}{3 N (N+1)} S_1
        \Biggr]
        +\textcolor{blue}{C_A} \Biggl[
                -\frac{8 P_5}{9 N^3 (N+1)^3}
\nonumber\\ &&
                +\frac{8 \big(
                        12+22 N+11 N^2\big)}{3 N^2 (N+1)} S_1
                -\frac{8 (2+N)}{N (N+1)} S_1^2
                +\frac{8 (2+N)}{N (N+1)} S_2
                +\frac{16 (2+N)}{N (N+1)} S_{-2}
        \Biggr]
\Biggr\}
\nonumber\\ &&
+\textcolor{blue}{C_F^2} \Biggl\{
        \frac{4 (2+N) (1+3 N) \big(
                -2-N+3 N^2+3 N^3\big)}{N^3 (N+1)^3}
        -\frac{8 (2+N) (1+3 N)}{N (N+1)^2} S_1
\nonumber\\ &&
        +\frac{8 (2+N)}{N (N+1)} S_1^2
        +\frac{8 (2+N)}{N (N+1)} S_2
\Biggr\},
\\
P_5 &=& 76 N^5+271 N^4+254 N^3+41 N^2+72 N+36,
\\
\gamma_{gg}^{(1)} &=&
\textcolor{blue}{C_F T_F N_F} \frac{8 P_8}{N^3 (1+N)^3}
+\textcolor{blue}{C_A T_F N_F} \Biggl\{
        \frac{32 P_6}{9 N^2 (1+N)^2}
        -\frac{160}{9} S_1
\Biggr\}
+\textcolor{blue}{C_A^2} \Biggl\{
        -\frac{4 P_9}{9 N^3 (1+N)^3}
\nonumber\\ &&
        +\Biggl(
                \frac{8 P_7}{9 N^2 (1+N)^2}
                -32 S_2
        \Biggr) S_1
        +\frac{64}{N (1+N)} S_2
        -16 S_3
        +\Biggl(
                \frac{64}{N (1+N)}
                -32 S_1
        \Biggr) S_{-2}
\nonumber\\ &&
        -16 S_{-3}
        +32 S_{-2,1}
\Biggr\},
\\
P_6 &=& 3 N^4+6 N^3+16 N^2+13 N-3,
\\
P_7 &=& 67 N^4+134 N^3+67 N^2+144 N+72,
\\
P_8 &=& N^6+3 N^5+5 N^4+N^3-8 N^2+2 N+4,
\\
P_9 &=& 48 N^6+144 N^5+469 N^4+698 N^3+7 N^2+258 N+144.
\end{eqnarray}
Here and in the following we only consider the non--singlet anomalous dimension $\gamma_{qq}^{(k),\rm NS} \equiv 
\gamma_{qq}^{(k),\rm NS,-}$. The anomalous dimensions agree with  results given in  
Refs.~\cite{Mertig:1995ny,SP_PS1,Vogt:2008yw,Moch:2014sna,Blumlein:2019zux}.

In the limit $N \rightarrow 1$, which can be performed using {\tt HarmonicSums}, we obtain
\begin{eqnarray}
\label{eq:gamGG1}
\gamma_{gg}^{(k)}(N = 1) &=& - 2 \beta_k,~~k \geq 0,
\end{eqnarray}
with 
\begin{eqnarray}
\beta_0 &=& \frac{11}{3} \textcolor{blue}{C_A} - \frac{4}{3} \textcolor{blue}{T_F N_F},
\\
\beta_1 &=& \frac{34}{3} \textcolor{blue}{C_A^2} 
- \frac{20}{3} \textcolor{blue}{C_A T_F N_F}
- 4 \textcolor{blue}{C_F T_F N_F},
\end{eqnarray}
cf.~\cite{Gross:1973id,Politzer:1973fx,Caswell:1974gg,Jones:1974mm}. Here, relation (\ref{eq:gamGG1}) is the 
consequence of the fact that the anomaly is maintained in its one--loop form, cf.~\cite{Larin:1993tq}.

As well the relation \cite{Moch:2014sna} 
\begin{eqnarray}
\label{eq:PSgq}
\gamma_{qq}^{(k),\rm PS}(N = 1) &=& - 4 \textcolor{blue}{T_F N_F}  
\gamma_{gq}^{(k-1)}(N = 1),~~k = 1,2. 
\end{eqnarray}
is verified. Furthermore,
\begin{eqnarray}
\label{eq:NSNone}
\gamma_{qq}^{(k),\rm NS}(N = 1) &=& 0,\\
\label{eq:qgNone}
\gamma_{qg}^{(k)}(N = 1) &=& 0,~~k \geq 0,
\end{eqnarray}
hold, where (\ref{eq:NSNone}) is implied by fermion number conservation. The other first moments are given by
\begin{eqnarray}
\gamma_{gq}^{(0)}(N = 1) &=&  6 \textcolor{blue}{C_F}\\
\gamma_{gq}^{(0)}(N = 1) &=&  -\frac{142}{3} \textcolor{blue}{C_F C_A} + 18 \textcolor{blue}{C_F^2} 
+ \frac{8}{3} \textcolor{blue}{C_F T_F N_F}.
\end{eqnarray}
\section{The contributions to the polarized three--loop anomalous dimensions \boldmath $\propto T_F$}
\label{sec:6}

\vspace*{1mm}
\noindent
We will first present the anomalous dimensions in Mellin--$N$ space for $N$ an odd integer. Later this will form the 
basis to 
transform to the splitting functions to $z$--space. We consider the polarized massive OMEs $A_{qq,Q}^{(2),\rm PS}, 
A_{Qq}^{(2),\rm PS}, A_{qg,Q}^{(2)}, A_{Qg}^{(2)}, A_{gq,Q}^{(2)}$ and $A_{gg,Q}^{(2)}$ for odd values of $N$ and 
derive the $T_F$-dependent
contributions to the anomalous dimensions from these quantities. After finally transforming from the Larin to the 
M--scheme one obtains:

All anomalous dimensions agree with the results of Ref.~\cite{Moch:2014sna}.

In the limit $N \rightarrow 1$  one obtains
\begin{eqnarray}
\hat{\gamma}_{gg}^{(2)}(N = 1) &=& - 2 \hat{\beta}_2,
\end{eqnarray}
with 
\begin{eqnarray}
\hat{\beta}_2 &=& 
-\frac{1415}{27} \textcolor{blue}{C_A^2 T_F}
-\frac{205}{9} \textcolor{blue}{C_A C_F T_F}
+ 2 \textcolor{blue}{C_F^2 T_F} 
+ \frac{158}{27} \textcolor{blue}{C_A T_F^2}
+ \frac{44}{9} \textcolor{blue}{C_F T_F^2} 
\nonumber\\ &&
+ \frac{316}{27} \textcolor{blue}{C_A T_F^2 N_F}
+ \frac{88}{9} \textcolor{blue}{C_F T_F^2 N_F}~,
\end{eqnarray}
cf.~\cite{Tarasov:1980au,Larin:1993tp}. Furthermore, Eqs.~(\ref{eq:PSgq},\ref{eq:NSNone},\ref{eq:qgNone}) hold analogously
and 
\begin{eqnarray}
\hat{\gamma}_{gq}^{(2)}(N = 1) &=& 
-\frac{164}{3} \textcolor{blue}{C_A C_F T_F} 
+ 214 \textcolor{blue}{C_F^2 T_F} 
+ \frac{104}{3} \textcolor{blue}{C_F T_F^2}
+ \frac{208}{3} \textcolor{blue}{C_F T_F^2 N_F} 
\nonumber\\ &&
+ 288 \textcolor{blue}{C_F T_F (C_A - C_F)} \zeta_3.
\end{eqnarray}
\section{The small \boldmath $z$ and large $N_F$ expansions}
\label{sec:7b}

\vspace*{1mm}
\noindent
In the polarized case the so--called leading singularity of the anomalous dimensions is situated at
$N=0$ for the complete singlet matrix and in the non--singlet case, unlike the unpolarized case 
\cite{Fadin:1975cb,Fadin:1998py}. In Refs.~\cite{Kirschner:1983di,Blumlein:1995jp,Bartels:1995iu}
predictions were made for the small--$z$ behaviour of the flavor non--singlet splitting functions
and in \cite{Bartels:1995iu,Blumlein:1996hb,Kiyo:1996si} and \cite{Blumlein:1996dd}
for the polarized non--singlet and singlet splitting functions in QCD and QED, although 
not being fully clear within which scheme, using so--called infrared evolution equations.

Up to two loops, it turned out that the most singular contributions around $N = 0$ for the 
anomalous  dimensions in the $\overline{\rm MS}$ scheme are predicted. This is not the case 
at three loop order, where only the diagonal elements agree\footnote{A corresponding deviation has also 
been observed for the sub--leading BFKL anomalous dimensions in the unpolarized case \cite{Fadin:1998py}.}. 
However, as shown in Ref.~\cite{Moch:2014sna}, if one interprets the predictions as physical  
evolution kernels, the `leading' terms of the corresponding anomalous dimensions agree.
In the future it has still to be checked whether or not BFKL predictions will hold to 
even higher orders, cf.~\cite{Blumlein:1998pp}.

An interesting question concerns the numerical effect of the sub--leading corrections to the 
leading terms $\propto 1/N^5$, see Refs.~\cite{Blumlein:1995jp,Blumlein:1996hb,Blumlein:1997em}.
One obtains
\begin{eqnarray}
\gamma_{qq}^{(2),\rm NS} &=& -\frac{64}{27 N^5} (29 - 132 N)  + O\left(\frac{1}{N^3}\right),
\\
\gamma_{qq}^{(2),\rm PS} &=& \frac{128}{3 N^5}(43 - 74 N) + O\left(\frac{1}{N^3}\right), 
\\ 
\gamma_{qg}^{(2)} &=& \frac{16}{N^5} (405-668 N) + O\left(\frac{1}{N^3}\right), 
\\
\gamma_{gq}^{(2)} &=& -\frac{64}{27 N^5}(2590-3907 N) + O\left(\frac{1}{N^3}\right), 
\\
\gamma_{gg}^{(2)} &=& -\frac{80}{3 N^5}(754 - 1217 N) + O\left(\frac{1}{N^3}\right).
\end{eqnarray}
The next sub--leading terms are more than canceling the leading terms, i.e. the leading terms 
alone yield basically nowhere dominant contributions in phenomenological applications numerically, despite the 
leading pole term of the complete calculation being correctly predicted. 

The finite renormalization between the Larin and the M--scheme does not affect the leading 
singular terms for the anomalous dimensions at $N=0$. 

We will now compare to predictions for the large $N_F$ terms given in 
Refs.~\cite{Gracey:1996ad,Bennett:1998sr}. Here we start with the flavor non--singlet case, 
\cite{Gracey:1996ad}, Eq.~(3.5), as the generating  function to understand the conventions 
used. The parameter $\mu$ is  $\mu= \hat{D}/2,~\hat{D}=4-2\hat{\ep},~\hat{\ep} \equiv (4/3) 
T_F \ep$ and $\ep$ denotes the usual dimensional 
parameter. The function $\eta_1^o$ is defined below in Eq.~(3.12) \cite{Gracey:1996ad}. The largest 
$N_F$ expansion coefficients to the non--singlet anomalous dimension are  now obtained as the 
coefficient $O(\ep^{k+1})$, with the leading order term at $O(\ep)$. We agree with the 
corresponding non--singlet predictions.

The generating function for the coefficients
\begin{eqnarray}
\label{eq:GR1}
\overline{\gamma}_{qq}^{(k)} -
\overline{\gamma}_{gq}^{(k)}
\frac{\overline{\gamma}_{qg}^{(0)}}{\overline{\gamma}_{gg}^{(0)}},~~k \geq 0
\end{eqnarray}
is given in Eq.~(5.3) \cite{Gracey:1996ad}, using the same definitions as in the non--singlet case 
above. Here $\overline{\gamma}_{ij}^{(k)}$ denotes the respective highest order term in $N_F$. 
Expanding as in the non--singlet case, we reproduce all terms containing harmonic sums but 
not the rational term in Eq.~(5.6) \cite{Gracey:1996ad} after rescaling by a factor $64 C_F 
T_F^2$ as suggested by the expansion. Taking the rescaled result (5.6) \cite{Gracey:1996ad} as 
it is and compare it to our three--loop result we obtain the difference
\begin{eqnarray}
-\textcolor{blue}{C_F T_F^2} \frac{2}{3} \frac{(N+2)(N^2-N-1)}{N^3(N+1)^3} = - 2 \beta_{0,Q} 
\hat{z}_{qq}^{(2) \rm PS}  
\end{eqnarray}
accounting for the normalization of the anomalous dimensions in the present paper. This suggests 
that the finite renormalization to the M--scheme for the pure singlet term in Eq.~(5.6) 
\cite{Gracey:1996ad} still has to be done, while 
$\overline{\gamma}_{qq}^{(2), \rm NS}$ and $\overline{\gamma}_{gq}^{(2)}$, as suggested by Eq.~(5.5) 
\cite{Gracey:1996ad}, are already in the M--scheme (note also a later discussion in Sect.~6 
\cite{Gracey:1996ad})\footnote{Unlike 
for the terms at two--loop order, which were presented individually, only the combination 
(\ref{eq:GR1}) has been presented in \cite{Gracey:1996ad}. Note that 
$\overline{\gamma}_{qq}^{(1) \rm PS}$ is scheme independent, while 
$\overline{\gamma}_{qq}^{(2) \rm PS}$ is not.}.

We compare now the large $N_F$ term for the combination
\begin{eqnarray}
\label{eq:GR2}
\overline{\gamma}_{gg}^{(2)} + 
\overline{\gamma}_{gq}^{(2)}  
\frac{\overline{\gamma}_{qg}^{(0)}}{\overline{\gamma}_{gg}^{(0)}} &=& 
-4 \textcolor{blue}{C_A T_F^2}  \Biggl[
       \frac{8 Q_1}{27 N^2 (1+N)^2} S_1
        +\frac{2 Q_2}{27 N^3 (1+N)^3}
\Biggr]
\nonumber\\ &&
+ 4\textcolor{blue}{C_F T_F^2}  \Biggl[
        -\frac{4 Q_3}{27 N^4 (1+N)^4}
        -\frac{64 (N-1) (N+2) \big(
                3+7 N+7 N^2\big)}{9 N^3 (1+N)^3} S_1
\nonumber\\ && \hspace*{1.7cm} 
        +\frac{64 (N-1) (N+2)}{3 N^2 (1+N)^2} S_1^2
\Biggr]
\\
Q_1 &=& 8 N^4+16 N^3-19 N^2-27 N+48,
\\
Q_2 &=& 87 N^6+261 N^5+249 N^4+63 N^3-76 N^2-64 N-96,
\\
Q_3 &=& 33 N^8+132 N^7+142 N^6-36 N^5-263 N^4-312 N^3+280 N^2
\nonumber\\ &&
+408 N+144,
\end{eqnarray}
\cite{Bennett:1998sr}, to which we agree in the M--scheme. 

We finally would like to add a remark by J. Gracey: In respect of the large $N_F$ computation of 
\cite{Gracey:1996ad}, to accommodate the finite renormalization constant associated with the Larin 
scheme in perturbation theory, whereby chirality is recovered in strictly four dimensions, a correction
had to be appended to the usual $D$--dimensional large $N_F$ critical exponent of the 
underlying operator. Such an additional piece is therefore by construction dependent on the procedure 
for handling $\gamma^5$. Hence the prediction for perturbative coefficients beyond the leading one of
\cite{Bennett:1998sr} may not tally with those for an alternative $\gamma^5$ definition as
appears to be the case here.
\section{Conclusions}
\label{sec:77}

\vspace*{1mm}
\noindent
We have calculated the contributions $\propto T_F$ to the polarized 3--loop anomalous dimension 
$\gamma_{ij}^{(2)}(N)$ and the associated splitting functions in a massive calculation, which is 
fully independent of the earlier computation in Ref.~\cite{Moch:2014sna}. We agree with the previous 
results. To have the opportunity to deal with propagator--based representations only, we used 
formal Taylor series representations in terms of an auxiliary parameter $x$ resumming the local operator 
insertions. As in the unpolarized case \cite{Ablinger:2017tan} before, we had to use the method of arbitrarily 
high moments \cite{Blumlein:2017dxp} to deal with potential elliptic contributions in the necessary deeper 
expansions in the dimensional parameter $\ep$ in the case of the OME $A_{Qg}^{(3)}$. 
In the method of high Mellin moments \cite{Blumlein:2017dxp} the moments  
are calculated recursively using the difference equation systems associated to the differential
equations given by the IBP relations. Individual master integrals are only calculated in terms of moments. 
In all other contributions, standard techniques, cf.~\cite{Blumlein:2018cms}, are used in the calculation of the 
master integrals.

The universality of the QCD anomalous dimension allows to compute them within various setups.
In the present calculation, they were obtained from the pole structure of massive polarized OMEs. 
The calculation of these OMEs is part of an ongoing project with the final goal to compute the 
massive polarized Wilson coefficients for deep--inelastic scattering in the region $Q^2 \gg m^2$.

The anomalous dimensions and splitting functions presented in this paper are also given in
{\tt Mathematica} format in ancillary files to this paper. The conversion to  {\tt maple} or {\tt 
FORM}-inputs \cite{FORM} is straightforward.

\appendix
\section{Feynman rules for local operator insertions}
\label{sec:A}

\begin{figure}[H]\centering
\includegraphics[width=0.25\linewidth]{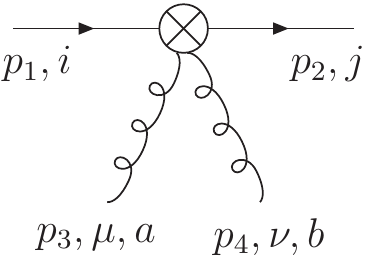} \hspace*{1cm}
\includegraphics[width=0.3\linewidth]{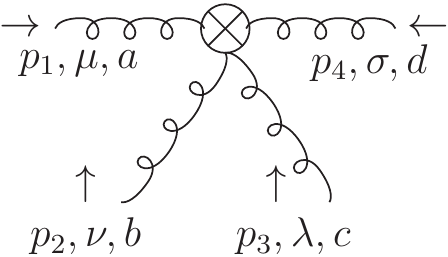}
\caption[]{\label{fig:FR} \sf The four-leg polarized local operator vertices.}
\end{figure}

\vspace*{1mm}
\noindent
Two of the Feynman rules given in Ref.~\cite{Mertig:1995ny} had to be corrected\footnote{We thank
J. Smith for a corresponding communication related to Ref.~\cite{Buza:1996xr} several years ago, 
with which we agree.}. They correspond to the vertices given in Figure~\ref{fig:FR}.
Here all momenta are ingoing. The quark-quark-gluon-gluon operator reads, with $N \in \mathbb{N}$ 
defining the Mellin  moment,
\begin{eqnarray}
O_{ab}^{\mu\nu}(p_1,p_2,p_3,p_4) &=& g^2 
\Delta^\mu
\Delta^\nu
\Delta \hspace*{-2.5mm}\slash \gamma_5 
\sum_{j=0}^{N-3} \sum_{l=j+1}^{N-2} (\Delta.p_2)^j (\Delta.p_1)^{N-l-2}
\nonumber\\ && \times
\Biggl[ 
(t_a t_b)_{kl} (\Delta.p_1 + \Delta.p_4)^{l-j-1}
+
(t_b t_a)_{kl} (\Delta.p_1 + \Delta.p_3)^{l-j-1} \Biggr],~N \geq 3.
\nonumber\\
\end{eqnarray}
The generators of the color group are denoted by $t_c$ and $g$ is  the strong coupling constant with
$a_s = g^2/(4\pi)^2$. The Feynman rule for the polarized four-gluon vertex reads
\begin{eqnarray}
O^{\mu\nu\rho\sigma}_{abcd}(p_1,p_2,p_3,p_4) &=& i g^2[1-(-1)^N][
  f_{abe} f_{cde} O^{\mu\nu\rho\sigma}(p_1,p_2,p_3,p_4)
\nonumber\\ &&
+ f_{ace} f_{bde} O^{\mu\rho\nu\sigma}(p_1,p_3,p_2,p_4)
- f_{ade} f_{bce} O^{\rho\nu\mu\sigma}(p_3,p_2,p_1,p_4)]
\nonumber\\ 
O^{\mu\nu\rho\sigma}(p,q,r,s) &=& 
(\ep^{\Delta\nu\rho\sigma} \Delta^\mu - \ep^{\Delta\mu\rho\sigma} \Delta^\nu)[\Delta.r + \Delta.s]^{N-2}
\nonumber\\ &&  
- \Delta^\rho (\ep^{\nu\sigma\Delta s} \Delta^\mu 
- \ep^{\mu\sigma\Delta s}\Delta^\nu) \sum_{i=1}^{N-3}[\Delta.r + \Delta.s]^i (\Delta.s)^{N-i-3}
\nonumber\\ &&
+ \Delta^\sigma(\ep^{\rho\nu\tau\Delta}\Delta^\mu-\ep^{\rho\mu r \Delta} \Delta^\nu)\sum_{i=0}^{N-3} 
[\Delta.r + \Delta.s]^{N-i-3} (\Delta.r)^i 
\nonumber\\ &&
- \Delta^\nu (\ep^{\mu\sigma\Delta p} \Delta^\rho - \ep^{\mu\rho\Delta p} \Delta^\sigma)
\sum_{i=0}^{N-3}[\Delta.r+\Delta.s]^{N-i-3}(-\Delta.p)^i
\nonumber\\ &&
+\Delta^\mu (\ep^{\nu\sigma\Delta q}\Delta^\rho - \ep^{\nu\rho\Delta q}\Delta^\sigma)
\sum_{i=0}^{N-3} [\Delta.r + \Delta.s]^{N-i-3} (-\Delta.q)^i
\nonumber\\ &&
+ \Delta^\nu \Delta^\rho (\ep^{\Delta\sigma p s} \Delta^\mu + \ep^{\mu\sigma \Delta s} \Delta.p)
\sum_{j=0}^{N-4} \sum_{i=0}^j (\Delta.p)^{N-j-4} [\Delta.p + \Delta.q)^{j-i} (-\Delta.s)^i
\nonumber\\ &&
- \Delta^\mu \Delta^\rho (\ep^{\Delta \sigma q s} \Delta^\nu + \ep^{\nu\sigma\Delta s} \Delta.q)
\sum_{j=0}^{N-4} \sum_{i=0}^j (\Delta.q)^{N-j-4} [\Delta.p + \Delta.q]^{j-i} (-\Delta.s)^i
\nonumber\\ &&
- \Delta^\nu \Delta^\sigma (\ep^{\Delta\mu p r} \Delta^\rho + \ep^{\mu\rho\Delta p} \Delta.r)
\sum_{j=0}^{N-4} \sum_{i=0}^j (\Delta.p)^{N-j-4}[\Delta.p + \Delta.q]^{j-i} (-\Delta.r)^i
\nonumber\\ &&
+ \Delta^\mu\Delta^\sigma (\ep^{\Delta\nu q r} \Delta^\rho + \ep^{\nu\rho\Delta p} \Delta.r)
\sum_{j=0}^{N-4} \sum_{i=0}^j (\Delta.q)^{m-j-4} (\Delta.p+\Delta.q)^{j-i} (-\Delta.r)^i.
\nonumber\\
\end{eqnarray}
Here the symbols $f_{abc}$ denote the QCD structure constants.

\noindent
Nevertheless the two--loop anomalous dimensions calculated in \cite{Mertig:1995ny} are correct.
\section{The splitting functions}
\label{sec:B}

\vspace*{1mm}
\noindent
The splitting functions are related to the anomalous dimensions by the Mellin transform
\begin{eqnarray}
\gamma_{ij}^{(k)}(N) = - \Mvec\left[P_{ij}^{(k)}(z)\right](N).
\end{eqnarray}
Since distribution--valued components are contributing, the Mellin transform is used as follows
\begin{eqnarray}
\Mvec\left[\left(f(z)\right)_+\right](N) &=& \int_0^1 dz \left(z^{N-1} - 1\right) f(z)
\\
\Mvec\left[g(z)\right](N) &=& \int_0^1 dz z^{N-1} g(z)
\\
\Mvec\left[\delta(1-z)\right](N) &=& 1.
\end{eqnarray}
In the following we present the complete polarized splitting functions to $O(\alpha_s^2)$ and the $T_F$-dependent
polarized splitting functions at $O(\alpha_s^3)$. Here $P_{qq}^{(2), \rm PS}$ and $P_{qg}^{(2)}$ are the complete 
splitting functions.

All splitting functions can be expressed by harmonic polylogarithms \cite{Remiddi:1999ew}, which are given by the iterated integrals
\begin{eqnarray}
\HA_{b,\vec{a}}(z) = \int_0^z dx f_b(x) \HA_{\vec{a}}(x),~~~~~\HA_\emptyset = 1, b, a_i \in \{0, 1, -1\},
\end{eqnarray}
over the alphabet
\begin{eqnarray}
f_c(z) \in \Biggl\{\frac{1}{z}, \frac{1}{1-z}, \frac{1}{1+z} \Biggr\}.
\end{eqnarray}
Again we reduce to the algebraic basis, cf.~\cite{Blumlein:2003gb}. As a shorthand notation we use $\HA_{\vec{a}}(z) 
\equiv \HA_{\vec{a}}$.
\subsection{The splitting functions to two--loop order}

\vspace*{1mm}
\noindent
The leading order splitting functions are given by
\begin{eqnarray}
P_{qq}^{(0)} &=& \textcolor{blue}{C_F} \Biggl\{
\frac{8}{(1 - z)}_+ -4 (1+z) + 6 \delta(1-z)
\Biggr\},
\\
P_{qg}^{(0)} &=& \textcolor{blue}{T_F N_F} \Biggl\{8 (-1 + 2 z)\Biggr\},
\\
P_{gq}^{(0)} &=& \textcolor{blue}{C_F} \Biggl\{4 (2 - z) \Biggr\},
\\
P_{gg}^{(0)} &=& \textcolor{blue}{C_A} \Biggl\{ \frac{8}{(1-z)}_+ + 8(1 - 2 z) \Biggr\}
+ 2 \left( \frac{11}{3} \textcolor{blue}{C_A} - \frac{4}{3} \textcolor{blue}{T_F N_F}
\right) \delta(1-z).
\end{eqnarray}
The next-to-leading order splitting functions read
\begin{eqnarray}
P_{qq}^{(1),\rm NS} &=& 
\Biggl\{
\Biggl[
\textcolor{blue}{C_F^2} \Biggl(
        -24 \HA_0
        +32 \HA_0 \HA_1
\Biggr)
+\textcolor{blue}{C_F C_A} \Biggl(
                \frac{8}{9} \big(
                        67
                        -18 \zeta_2
                \big)
                +\frac{88}{3} \HA_0
                +8 \HA_0^2
        \Biggr)
\nonumber\\ &&
        -\frac{32}{9} \textcolor{blue}{C_F T_F N_F} \big(
                5+3 \HA_0\big)
\Biggr] \frac{1}{1-z}\Biggr\}_+
+
\Biggl[
        -\frac{4}{3} \textcolor{blue}{C_F T_F N_F} 
                +\textcolor{blue}{C_A C_F} \Biggl(
                        \frac{17}{3}
                        -8 \zeta_3
                \Biggr)
\nonumber\\ &&
        +\textcolor{blue}{C_F^2} \big(
                3
                +16 \zeta_3
        \big)
\Biggr] \delta(1-z)
+ \textcolor{blue}{C_F T_F N_F} \Biggl(
                \frac{16}{9} (-1+11 z)
                +\frac{16}{3} (1+z) \HA_0
        \Biggr)
\nonumber\\ && 
        +\textcolor{blue}{C_A C_F} \Biggl(
                \frac{4}{9} (89-223 z)
                +\frac{4}{3} (1+z) \HA_0
                -\frac{16 \big(
                        1+z^2\big)}{1+z} \HA_{-1} \HA_0
                -\frac{8 z }{1+z} \HA_0^2
\nonumber\\ &&                
 +\frac{16 \big(
                        1+z^2\big)}{1+z} \HA_{0,-1}
                +\frac{16 z}{1+z} \zeta_2
        \Biggr)
+\textcolor{blue}{C_F^2} \Biggl(
        72 (-1+z)
        -16 (1+2 z) \HA_0
\nonumber\\ &&    
     +\frac{32 \big(
                1+z^2\big)}{1+z} [\HA_{-1} \HA_0 - \HA_{0,-1}]
        -\frac{4 \big(
                3+2 z+3 z^2\big)}{1+z} \HA_0^2
        -16 (1+z) \HA_0 \HA_1
\nonumber\\ &&        
 +\frac{16 \big(
                1+z^2\big)}{1+z} \zeta_2
\Biggr),
\\
P_{qq}^{(1),\rm PS} &=& \textcolor{blue}{C_F T_F N_F} \Biggr\{
         16 (1-z)
        -16 (1-3 z) \HA_0
        -16 (1+z) \HA_0^2
\Biggr\},
\\
P_{qg}^{(1)} &=& \textcolor{blue}{C_A T_F N_F}
\Biggl\{
        16 (12-11 z)
        +16 (1+8 z) \HA_0
        -32 (1+2 z) \HA_{-1} \HA_0
        -16 (1+2 z) \HA_0^2
\nonumber\\ &&         
+64 (1-z) \HA_1
        +16 (1-2 z) \HA_1^2
        +32 (1+2 z) \HA_{0,-1}
        -32 \zeta_2
\Biggr\}
\nonumber\\ && 
+ \textcolor{blue}{C_F T_F N_F} \Biggl\{
        8 (-22+27 z)
        -72 \HA_0
        -8 (1-2 z) \HA_0^2
        -\Biggl(
                64 (1-z)
 +32 (1-2 z) \HA_0
        \Biggr) \HA_1
\nonumber\\ &&                
        +16 (-1+2 z) \HA_1^2
        -32 (-1+2 z) \zeta_2
\Biggr\},
\\
P_{gq}^{(1)} &=& \textcolor{blue}{C_F T_F N_F}
\Biggl\{
                -\frac{32}{9} (4+z)
                +\frac{32}{3} (2-z) \HA_1
        \Biggr\}
        +\textcolor{blue}{C_A C_F} \Biggl\{
                \frac{8}{9} (41+35 z)
                +8 (4-13 z) \HA_0
\nonumber\\ &&
                +16 (2+z) \HA_{-1} \HA_0
                +8 (2+z) \HA_0^2
                -\Biggl(
                        \frac{8}{3} (10+z)
                        +16 (-2+z) \HA_0
                \Biggr) \HA_1
                -8 (-2+z) \HA_1^2
\nonumber\\ && 
                -16 (2+z) \HA_{0,-1}
                +16 z \zeta_2
\Biggr\}
+\textcolor{blue}{C_F^2} \Biggl\{
        4 (-17+8 z)
        +4 (-4+z) \HA_0
        -4 (-2+z) \HA_0^2
\nonumber\\ &&
        +8 (2+z) \HA_1
        +8 (-2+z) \HA_1^2
\Biggr\},
\\
P_{gg}^{(1)} &=& 
\Biggl\{\Biggl[
- \textcolor{blue}{C_A T_F N_F} \frac{160}{9} 
+\textcolor{blue}{C_A^2} \Biggl(
        -\frac{8}{9} \big(
                -67
                +18 \zeta_2
        \big)
        +8 \HA_0^2
        +32 \HA_0 \HA_1 \Biggr)
\Biggr] \frac{1}{1-z}\Biggr\}_+
\nonumber\\ &&
+ \Biggl[
        -\frac{32}{3}\textcolor{blue}{C_A T_F N_F}
        - 8 \textcolor{blue}{C_F T_F N_F}
        +  \textcolor{blue}{C_A^2}
               \left(8 \zeta_3 +\frac{64}{3}\right)
\Biggr] \delta(1-z)
\nonumber\\ && 
+ \textcolor{blue}{C_A T_F N_F} \Biggl(
        \frac{32}{9} (-14+19 z)
        -\frac{32}{3} (1+z) H_0(z)
\Biggr)
+ \textcolor{blue}{C_F T_F N_F} \Biggl(
        80 (-1+z)
\nonumber\\ && 
        +16 (-5+z) \HA_0
        -16 (1+z) \HA_0^2
\Biggr)
+\textcolor{blue}{C_A^2} \Biggl(
        -\frac{4}{9} (37+97 z)
        -\frac{8}{3} (-29+67 z) \HA_0
\nonumber\\ && 
+\frac{32 \big(
                2+3 z+2 z^2\big)}{1+z} [\HA_{-1} \HA_0 -  \HA_{0,-1}]
        +\frac{8 (3+4 z)}{1+z} \HA_0^2
        +32 (1-2 z) \HA_0 \HA_1
\nonumber\\ && 
        +\frac{16 (1+2 z)^2 }{1+z} \zeta_2
\Biggr).
\end{eqnarray}

\subsection{The contributions \boldmath $\propto T_F$  at three--loop order}

\vspace*{1mm}
\noindent
We obtain the following contributions  $\propto T_F$ to the next-to-next-to-leading order polarized splitting 
functions


\vspace*{5mm}
\noindent
{\bf Acknowledgment.} We thank J.~Gracey, F.~Herren, S.~Moch, D.~St\"ockinger, A.~Vogt and S.~Weinzierl for 
discussions. This work has been funded in part by EU TMR network SAGEX agreement No. 764850
(Marie Sk\l{}odowska-Curie) and COST action CA16201: Unraveling new physics at
the LHC through the precision frontier. The Feynman diagrams were drawn using {\tt Axodraw} \cite{Vermaseren:1994je}.

\end{document}